# Noise analysis, error estimates, and Gamma Radiation Measurement for limited detector computerized tomography application

Kajal Kumari[1$] , Mayank Goswami[1#],

[1]Divyadrishti Laboratory, Department of Physics, IIT Roorkee, Roorkee, India

Department of Physics, IIT Roorkee, Roorkee, India

[#]mayank.goswami@ph.iitr.ac.in

[$]kkumari@ph.iitr.ac.in and [$]kajalkumari7867@gmail.com

## Abstract

Computed Tomography is one of the efficient and vital modalities of non-destructive techniques (NDT). Various factors influence the CT reconstruction result, including limited projection data, detector electronics optimization, background noise, detection noise, discretized nature of projection data, and many more. Radiation hardening and other aging factors that affect the operational settings may require recalibration of electronics parameters. Two well-known exercises are utilized with the motivation to improve reliability and accuracy in inverse recovery. The first exercise brute-forces an optimal candidate from the set of calibration methods for minimum error in inverse recovery. The second exercise, Kanpur Theorem-1 (KT-1) examines if optimal calibration sets electronics to impart minimum noise. The mutual conformity between statistics-derived CLT and Riemann integral transform-based KT-1 is shown first time using gamma radiation measurement. The analysis shows that measurement data with normal distribution inflicts the least noise in inverse recovery.

## 1. Introduction

A standard CT has a radiation source (such as X-ray Tube, Gamma Ray source, Ultrasounds Transducer, etc.), detector array, and their corresponding electronics. The quality of CT results is influenced by an inherent error that can occur for a variety of reasons, including the data collection process, reconstruction algorithm, mismatch or fluctuation in the detector's electronics settings, and the discretized nature of projection data due to the uncertainty of radiation emergence in radioactive sources[1]. The measurement requires synchronization between the control electronics of electromechanical assembly and the detector electronics[2]. Sophisticated automation technologies are well developed[3]. Noise due to changes in the environment can be shielded. The unavoidable/inherent measurement noise gets amplified if the detector electronic settings are not set optimally. Generally, the manufacturer provides these settings. Normal operational aging of the CT setup may change these optimal settings[4], [5]. Moreover, ionizing radiation modality suffers against radiation hardening. The detection electronics is complex but easily controllable in Gamma Radiation Modality. It is chosen to address and highlight the related issue of electronic noise associated with CT projection data at non-optimized electronic settings. Sensitivity analysis of several calibration methods for optimal detector electronics setting is presented[6].

Limited Detector Computed Tomography (LDCT) systems may be the only option for non-invasive investigation sometimes. Scanning object vs. total detector array size, limited economical budget, and/or space limitation around the scanning area are few constraints. The mathematical modeling of LDCT falls under the sparse data inverse problem resulting into ill-posed underdetermined system of equations[7], [8]. The quality (reliability and accuracy) of the reconstruction may get affected by factors imparting inherent error such as measurement noise in projection data providing a random result. Verification of accuracy requires destructive analysis ultimately ill serving the purpose of non-destructive analysis. An already established mathematical framework named 'Kanpur Theorem' [9] gives the qualitative estimation of the noise level in the measured data as reliability check in process of recovery.

Motivation

Two exercises are used in this study to improve the accuracy and reliability of inverse recovery. In the first exercise, the optimal setting is obtained, using brute-force combination of the different values of electronics parameters, which yields a reconstruction image with least root mean square error (RMSE). This excersice sculpts the probability distribution of signal profile towards normal distribution improving the statistics of measuring true signal and supressing noise prior to inverse recovery process. Later, the Kanpur Theorem is used, which employs a variety of filter functions to suppress undesirable signals based on filter selection, resulting in less contaminated measurement data.

2. Theory and Experiment

We have carried out an experiment to measure radiation counts at a set of different combinations of electronic settings of the scintillation detector. Gamma CT experiment is performed at the settings calculated by five methods of calibration: Central Limit Theorem-1 (CLT-1), CLT-2, pulse width saturation method (PWS), and true counts measurement method, gross method, and one at the setting, provided by the manufacture. Two different schemes have been used in this work to find out the less corrupted data. In the first one, the best method and the best projection data are obtained by comparing the reconstructed image's standard deviation error, i.e., RMSE[10]. The second one is Kanpur Theorem, which is briefly explained in the next sub-section 'Kanpur Theorem'.

Kanpur Theorem

Kanpur Theorem-1 is used to measure the level of noise in the measurement data. This fact motivates an indirect reconstruction error when the cross-sectional distribution of real objects is unknown, unlike the simulated cyber phantoms. Sobolev space-based KT-1 is used, when the user has projection data, to estimate the level of noise in projection data while KT-2 is used to characterize the CT images in the absence of projection data. For a given data set, the sharpness corresponding to the maximum grayscale value of the reconstructed image can be used as an error indicator that is related to the Delta Function Response of the filters chosen to reconstruct the images. A plot between the reciprocal of the maximum grayscale value of the reconstructed image "$1/N_{max}$" and the second derivative of filter function $W''(0)$ is used to check the correctness of tomographic data. A linear relationship shows that the projection data is noise-free or has a low level of noise. Different filter functions have been incorporated into the reconstruction procedure to improve the reconstruction image result. The value of the second derivative of the filter function is shown in table1.

Table 1: Tomographic filters and their second derivative values used in this study.

| Code | Class   | Nature | $W''(0)$ |
|------|---------|--------|----------|
| h99  | Hamming | Sharp  | 0.001    |
| h91  | Hamming | Sharp  | 0.083    |
| h75  | Hamming | Medium | 0.250    |
| h54  | Hamming | Smooth | 0.460    |
| h50  | Hamming | Smooth | 0.500    |

Experimental Details

The experiment comprises an isotopic Cs-137 gamma radioactive source of activity 1.5$\mu$Ci and scintillation detectors. The radiation detectors used in our work are five NaI(Tl) scintillator crystals coupled with a Photomultiplier tube whose output is delivered to a scintillation counting system i.e. Single Channel Analyser (SCA) and one radiation detection module comprised of scintillator crystal CsI(Tl) and multi-pixel photon counter to detect gamma radiation[6]. We used NaI(Tl) scintillation crystal for the detection of gamma radiation since it yields a good photon detection efficiency, is cost-effective, and is suitable to use at room temperature[11],[12]. Later one is used in our work as true counts measurement detector in one of the proposed methods of calibration.

The modalities of the scintillator detector give control over the electronics parameters. Button/knob are available to change the high voltage value at photomultiplier tube (PMT), amplifier gain, lower limit discrimination (LLD) value, and SCA require the average time input value to count the gamma radiation. The above four parameters are changed independently in apt ranges and counts are measured at these settings.

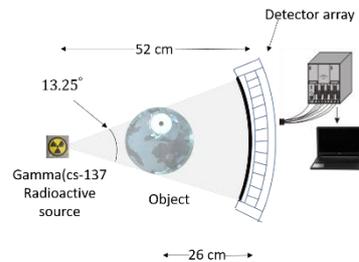
Fig. 1: Gamma CT geometry

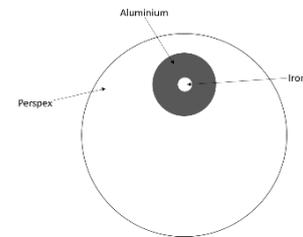
Fig. 2: Distribution of materials in phantom

Different methods of detector calibration that had been proposed in our previous work are utilized in the current work[6]. Optimal settings of electronics parameters of scintillator detectors are calculated corresponding to each proposed calibration method.

Gamma CT experiment is performed at different settings calculated by five different calibration methods and one at the setting provided by the manufacturers at the time of installation of the scintillator detector. The experimental setup of gamma CT is shown in fig. 1. A known cylindrical-shaped phantom is used for scanning purposes.

The phantom is made up of Perspex material of size 12 cm in diameter. Two holes with diameters of 3.8 cm and 0.8 cm are drilled, later filled with aluminium ($\mu{\sim}0.211$,) and iron ($\mu{\sim}0.544$) material. The distribution of materials in phantom is shown in figure 2. Fan beam geometry is used to collect the projection data in a gamma CT experiment of fan beam angle $13.25°$ in which five NaI(Tl) scintillator detectors are placed in an arc. Source to detector and object to detector distance is kept at 52cm and 26cm respectively. The projection data is collected in 360 rotations with an angle increment of $10°$.

3. Results and Discussion

Reconstruction of Experimental Data

Kanpur Theorem gives the error estimation in convolution backprojection algorithm (CBP) reconstructions to the choice of the filter functions generally selected by users[13]. Fanbeam geometry mode has been used while performing the Gamma CT experiment. To utilize the Kanpur Theorem as an indicator of error estimation, first, we have converted our fan-beam projection data into the parallel beam and then MATLAB® inbuilt function of CBP named 'iradon' is used to reconstruct the projection data of the scanned object. Apt values of input information such as geometry information, number of detectors, number of rotations, and type of filter functions are given.

## Analysis

The cross-section distribution of the phantom has been reconstructed at six different settings using different filter functions listed in table 1. A graph has been plotted between $1/N_{max}$ and $W''(0)$. KT-1 signature of the experimental data is shown in figure 3. The linear model is used to fit the experimental data at different filter functions. The root means square error of the linear fitting ($RMSE_{KT1}$) is used as a parameter to check the linear behavior of the KT-1 signature. $RMSE_{KT-1}$ for different projection data obtained at different electronics settings is shown in figure 4. It can be seen clearly from figure 4 that minimum $RMSE_{KT1}$ has been obtained for projection data that was collected at electronics settings calculated by CLT-1 method. It indicates that the CLT method provides the electronic settings at which minimum noise is obtained. Most researchers used the parameter to check the reconstruction error in projection data for the known object, which is the deviation of the reconstructed image from the simulated cyber phantom pixel to pixel.

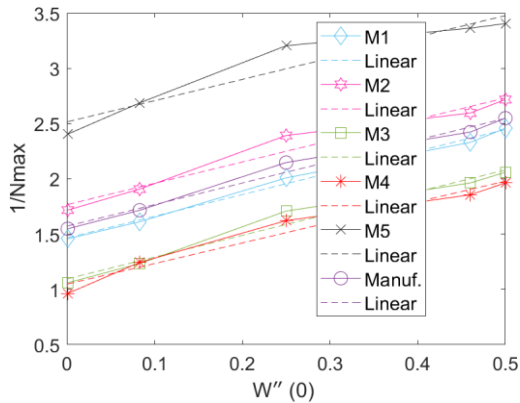

Fig. 3: KT-1 Signature

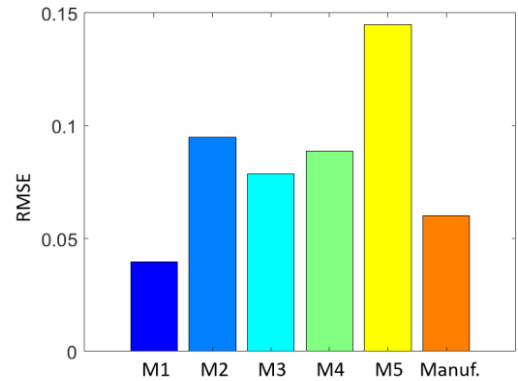

Fig. 4: $RMSE_{KT1}$ of KT-1 signature

We have plotted the bar graph of the Root mean square error $RMSE_{CT}$ or deviation of the reconstructed image w.r.t. cyber phantom obtained at six different electronic settings. This $RMSE_{CT}$ is shown in figure 5. It shows that the reconstruction result obtained at electronics settings, which are calculated by CLT-2 method, has more similarity with the simulated cyber phantom. $RMSE_{CT}$ for CLT-2 method and manufactured settings are very close which shows electronics settings calculated by CLT-2 method and the electronics settings provided by the manufacturer yield the CT images having good similarity with cyber phantom as compared to results obtained at other settings.

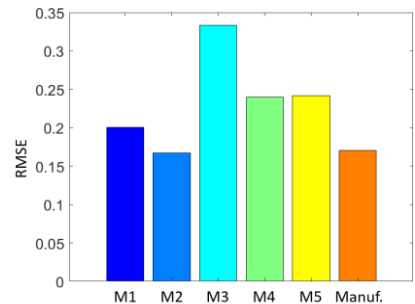

Fig. 5: $RMSE_{CT}$ at different electronics settings.

If we see the noise level in projection data obtained at different settings, it has been observed that after CLT-1 method, manufacturer settings preserve minimum noise in projection data. $RMSE_{KT1}$ shows CLT-1 while the similarity-based parameter $RMSE_{CT}$ shows CLT-2 methods is a best method, if we check their definitions then it is found that both exercises preserve their respective version of CLT. The reconstructed image obtained at electronic settings calculated by CLT-1, CLT-2, and manufactured settings for different filter functions are shown in figure 6. Figure 6(b) and 6(c) shows that $RMSE_{CT}$ is not a reliable parameter to consider. $RMSE_{CT-M2-h54}$ for reconstruction from M2 setting is less than $RMSE_{CT-Manuf-h54}$ from Manufacturer setting but visual comparison between these two figure reveals inverse of this conclusion. However, we chose that final decision should be taken on human perception.

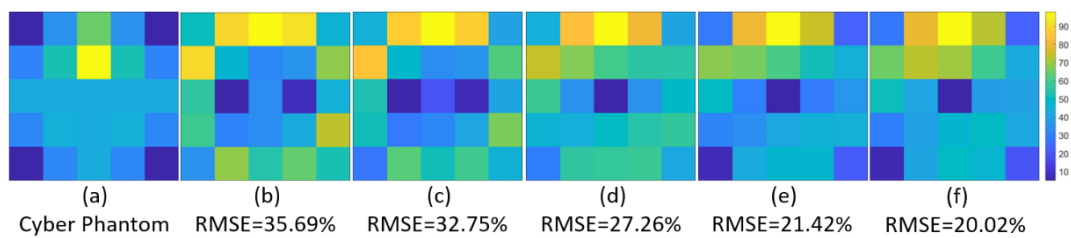

Fig. 6(a): (a) Simulated cyber phantom, reconstructed image at M1 setting with (b) h99 (c) h91 (d) h75 (e) h54 (f) h50 Hamming class Filters.

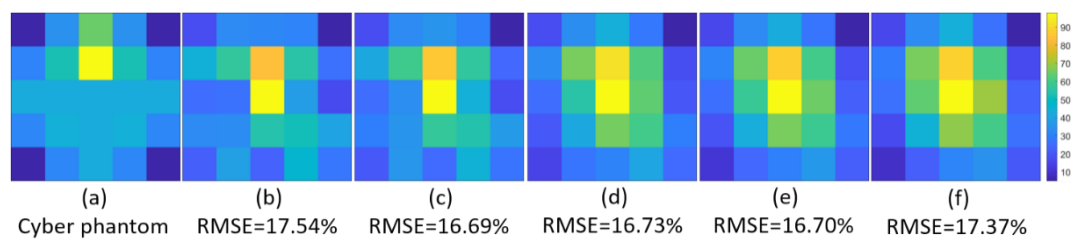

Fig. 6(b): (a) Simulated cyber phantom, reconstructed image at M2 setting with (b) h99 (c) h91 (d) h75 (e) h54 (f) h50 Hamming class Filters.

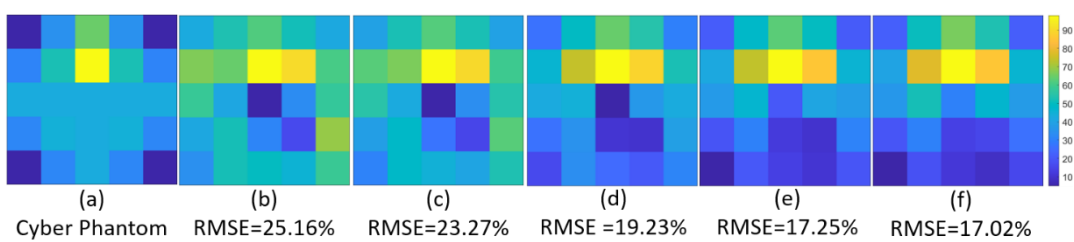

Fig. 6(c): (a) Simulated cyber phantom, reconstructed image at manufacturer setting with (b) h99 (c) h91 (d) h75 (e) h54 (f) h50 Hamming class Filters.

### 4. Conclusion

A major finding of this study is that the measurement noise associated with detector electronics may have several components. The one that is getting detected by KT1 is same that is getting suppressed by using the detector electronics paramteres estimated via CLT. However, the other components which are effecting the recovery algorithm thus visual quality of reconstruction is not being detected by both of these tools. Performance evaluation parameter other than $RMSE_{CT}$ must be used in CT analysis.